# New Developments in Precision LHC Theory: QED⊗QCD Exponentiation, Shower/ME Matching, IR-Improved DGLAP-CS Theory and Implications for UV Finite Quantum Gravity

**B.F.L. Ward[1]**
*Department of Physics, Baylor University*
*Waco, Texas, USA*
*E-mail: BFL_Ward@baylor.edu*

**S.A. Yost**
*Department of Physics, Princeton University*
*Princeton, New Jersey, USA*
*E-mail: SYOST@princeton.edu*

**Abstract**

We present the recent developments in exact amplitude-based resummation methods for non-Abelian gauge theories as they relate to precision LHC physics. We discuss QED⊗QCD exponentiation, shower/ME matching, IR-improved DGLAP-CS theory and implications, as developed by one of us (BFLW), for a UV finite theory of quantum general relativity.

---

[1] Speaker

## 1.     Introduction

The impending start-up of the LHC makes urgent the calculation, in a practical way, precision predictions, at or better than 1% precision, for the Standard Model and certain beyond the Standard Model processes relevant in LHC scenarios. In direct terms, exact results are needed at the $O(\alpha_s^2)L^n$, $O(\alpha_s\alpha)L^{n'}$, $O(\alpha^2)L^{n''}$, where $n, n' = 0, 1, 2$, and $n'' = 1, 2$, with consistent soft gluon and soft photon resummation in the presence of parton showers. Here, $L$ represents the big logs that characterize the specific process under study, some of which reflect for example the realistic effects of experimental cuts. In what follows, we present the recent progress in the QED⊗QCD resummation platform which we have developed [1] to address such precision LHC physics.

We stress here that any platform which purports to realize 1% or better precision in QCD in the presence of the now clear large effects from electroweak corrections for LHC processes [2] must have a clear proof of control over the error budget for both physical and technical precision components. Our platform, based ultimately as it is on multiple gluon and multiple photon MC methods, affords us such control, as we explain more completely presently.

We proceed as follows with the discussion. In the next Section, we review the exact amplitude-based resummation theory for QCD. Section 3 presents its extension to QED⊗QCD and quantum gravity. Section 4 presents applications and recent developments for the LHC physics and Section 5 discusses applications of general theoretical framework by one of us (BFLW) to the final state of Hawking radiation[3] in quantum gravity.

## 2.     Exact Amplitude-Based Resummation for QCD

In the case of QED, it has been shown in Refs.[4] that the amplitude-based resummation calculus of Yennie, Frautschi and Suura (YFS) [5] allows practical, precision Monte Carlo simulation of higher order radiative processes on an event-by-event basis with realistic multiple $n(\gamma)$ radiation in which infrared (IR) singularities are cancelled to all orders in $\alpha$. This amplitude-based approach to resummation theory in quantum field theory has been extended to QCD in Refs. [6]. We now review this extension.

The authors in Refs. [6] have shown that, for the process such as $q(p_1)+q'(q_1) \rightarrow q''(p_2)q'''(q_2)+n(G)$, we have the differential cross section, derived from extending the YFS theory to QCD,

$$d\hat{\sigma}_{\exp} = \sum_n d\hat{\sigma}_n$$

$$= e^{SUM_{IR}(QCD)} \sum_{n=0}^{\infty} \frac{1}{n!} \int \prod_{j=1}^{n} \frac{d^3 k_j}{k_j} \int \frac{d^4 y}{(2\pi)^4} e^{iy(p_1+q_1-p_2-q_2-\sum_j k_j)+D_{QCD}}$$

$$* \tilde{\bar{\beta}}_n(k_1,\ldots,k_n) \frac{d^3 p_2}{p_2^0} \frac{d^3 q_2}{q_2^0} \qquad (1)$$

where infrared functions $SUM_{IR}(QCD)$ and $D_{QCD}$ and hard gluon residuals $\tilde{\bar{\beta}}_n$ are defined in Refs.[6] – the latter are free of IR singularities to all orders in $\alpha_s$.

The exponent in (1) corresponds to the *N=1* term in the exponent in Gatheral's non-Abelian eikonal formula [7] wherein everything that does not eikonalize and exponentiate is dropped – our result in (1) is exact. There is no problem at all to include all of Gatheral's exponent in (1), as it would amount to a corresponding change in the residuals $\widetilde{\bar{\beta}}_n$.

The result in (1) has to be DGLAP-CS synthesized[6,8] via the QCD factorization theorem at a scale $\mu$ to be convoluted with parton densities at such scale. In general, this means that the infrared functions $SUM_{IR}(QCD)$ and $D_{QCD}$, as explained in [6,8], are replaced with the DGLAP-CS synthesized functions $SUM_{IR}^{nls}(QCD)$ and $D^{nls}{}_{QCD}$, where we note

$$SUM_{IR}^{(nls)}(QCD) = 2\alpha_s \operatorname{Re} B_{QCD}^{(nls)} + 2\alpha_s \widetilde{B}_{QCD}^{(nls)}, \quad D_{QCD}^{(nls)} = D_{QCD}(\widetilde{S}_{QCD}^{(nls)}),$$

with the infrared functions $B_{QCD}^{(nls)}, \widetilde{B}_{QCD}^{(nls)}, \widetilde{S}_{QCD}^{(nls)}$ and $D_{QCD}$ defined in [6].

## 3.  Extension to QED⊗QCD and Quantum Gravity

In Refs.[1], we have extended the master formula in (1) to treat the simultaneous amplitude resummation of QED and QCD together, in view of the relatively large electroweak effects that obtain in LHC physics scenarios in the *1%* QCD regime, and one of us (BFLW)[9] has extended (1) to treat resummation of quantum gravity in view of the similarity between the soft gluon and soft graviton limits in quantum field theory, as explained in Refs.[9]. We now review these results.

For the case of the QED⊗QCD resummation, we show in Ref. [1] that the master formula for the resummed cross section is

$$d\hat{\sigma}_{\exp} = \sum_{n,m} d\hat{\sigma}_{n,m}$$
$$= e^{SUM_{IR}(QCED)} \sum_{n=0}^{\infty} \frac{1}{n!m!} \int \prod_{j_1=1}^{n} \frac{d^3 k_{j_1}}{k_{j_1}} \prod_{j_2}^{m} \frac{d^3 k'_{j_2}}{k'_{j_2}} \int \frac{d^4 y}{(2\pi)^4} e^{iy(p_1+q_1-p_2-q_2-\sum_{j_1} k_{j_1} - \sum_{j_2} k'_{j_2}) + D_{QCED}}$$
$$* \widetilde{\bar{\beta}}_{n,m}(k_1,\ldots,k_n;k'_1,\ldots,k'_m) \frac{d^3 p_2}{p_2^0} \frac{d^3 q_2}{q_2^0} \quad (2)$$

where now we have the IR functions

$$SUM_{IR}^{(nls)}(QCED) = 2\alpha_s \operatorname{Re} B_{QCED}^{(nls)} + 2\alpha_s \widetilde{B}_{QCED}^{(nls)}, \quad D_{QCED}^{(nls)} = D_{QCED}(\widetilde{S}_{QCED}^{(nls)}),$$

with the identifications $A_{QCED} = A_{QCD} + A_{QED}$ for $A = B^{(nls)}, \widetilde{B}^{(nls)}, D^{(nls)}$, and $\widetilde{S}^{(nls)}$. Here, we have introduced the QED IR functions familiar from the original work in Ref. [5]. The new residuals $\widetilde{\bar{\beta}}_{n,m}$ are for *n* hard gluons and *m* hard photons and are free of IR singularities to all orders in both $\alpha_s$ and $\alpha$. The relative sizes of the two latter coupling parameters cause the QCD emission to occur an order of magnitude earlier than does the respective QED emission in the soft regime[1]. After DGLAP-CS synthesization, the residuals are then truly perturbative and the leading term $\widetilde{\bar{\beta}}_{0,0}^{(0,0)}$ gives us a good estimate of the size of the effects we study.

Applying the master formula (1) to quantum general relativity one finds[9] that the propagator for a massive scalar field has the representation

$$i\Delta'_F(k)|_{\text{resummed}} = \frac{ie^{B''_g(k)}}{k^2 - m^2 - \Sigma'_s(k) + i\varepsilon} \qquad (3)$$

for

$$B''_g(k) = -2i\kappa^2 k^4 \int \frac{d^4\ell}{16\pi^4} \frac{1}{\ell^2 - \lambda^2 + i\varepsilon} \frac{1}{(\ell^2 + 2\ell k + \Delta + i\varepsilon)^2} \qquad (4)$$

with $\Delta = k^2 - m^2$. Here, $\lambda$ is an IR regulator mass—it is not a parameter in the Lagrangian so that it does not affect the gauge invariance of Einstein's theory. One notes that $\Sigma'_s$ starts in $O(\kappa^2)$ so that it may me dropped in one-loop results. One notes further that in the deep UV one has

$$B''_g = \frac{\kappa^2 |k^2|}{8\pi^2} \ln\left(\frac{m^2}{m^2 + |k^2|}\right) \qquad (5)$$

so that loop integrals are now all UV finite[9].

## 4.0 QED⊗QCD threshold corrections, shower/ME matching, and IRI—DGLAP-CS theory at LHC

We apply the master formula (2) to single Z production with leptonic decay to focus on the ISR alone. See also Refs.[10] for exact $O(\alpha)$ results and Refs.[11,12] for exact $O(\alpha_s, \alpha_s^2)$ results. For the basic formula

$$d\sigma_{\exp}(pp \to V + X \to \ell\bar{\ell}' + X') = \sum_{i,j} \int dx_1 dx_2 F_i(x_1) F_j(x_2) d\hat{\sigma}_{\exp}(x_1 x_2 s) \qquad (6)$$

we use the result in (2) with parton densities $\{F_i\}$ from [13], where here we employ semi-analytical methods at the $\tilde{\bar{\beta}}^{(0,0)}_{0,0}$ - level. A MC realization will appear elsewhere[14]. We further note that we do not intend to replace Herwig/Pythia[15,16]. We intend to match the latter to our exact YFS-style calculus[1]: we can do this by the $p_T$-matching prescription or the shower - subtracted residuals, $\tilde{\bar{\beta}}_{n,m} \to \hat{\tilde{\bar{\beta}}}_{n,m}$, as they are defined in Ref.[1]—see also Ref.[17]. Each of these approaches can be systematically improved with exact results order by order in $(\alpha, \alpha_s)$ with exact phase space. Recent alternative parton evolution algorithms as given in Refs. [18] are readily accommodated in our approach. Proceeding now as indicated, we compute the ratio $r_{\exp} = \sigma_{\exp} / \sigma_{\text{Born}}$, with and without QED, to get the results ( we do not use the narrow resonance approximation)

$$r_{\exp} = \begin{cases} 1.1901 & \text{QCED} \equiv \text{QED} \otimes \text{QCD} & \text{LHC} \\ 1.1872 & \text{QCD} & \text{LHC} \\ 1.1911 & \text{QCED} & \text{Tevatron} \\ 1.1879 & \text{QCD} & \text{Tevatron} \end{cases} \qquad (7)$$

which show that QED is at the 0.3% level at both LHC and FNAL. The results are stable against scale variations and agree with those in Refs.[10-12]. The results are similar in size to the structure function results in Ref.[19] and DGLAP-CS synthesization has not compromised the normalization.

The results (7) then lead one to ISR exponentiation of the kernels in DGLAP-CS themselves in order to achieve reliable MC implementation of (2)[8]. One finds[8] that (1) applied to the parton branching processes allows improvement of the respective kernels to

$$P_{qq}(z) = C_F F_{YFS}(\gamma_q) e^{\frac{1}{2}\delta_q} \left[ \frac{1+z^2}{1-z}(1-z)^{\gamma_q} - f_q(\gamma_q)\delta(1-z) \right],$$

$$P_{Gq}(z) = P_{qq}(1-z), \ z<1,$$

$$P_{GG}(z) = 2C_G F_{YFS}(\gamma_G) e^{\frac{1}{2}\delta_G} \{(1-z)z^{\gamma_G-1} + z(1-z)^{\gamma_G-1} + (z^{1+\gamma_G}(1-z) + z(1-z)^{1+\gamma_G})/2 - f_G(\gamma_G)\delta(1-z)\},$$

$$P_{qG}(z) = F_{YFS}(\gamma_G) e^{\frac{1}{2}\delta_G} \frac{1}{2}\{z^2(1-z)^{\gamma_G} + (1-z)^2 z^{\gamma_G}\} \qquad (9)$$

where the IR exponents $\gamma_j$ and normalizations $f_j$, $j=q, G$, are given in Refs. [8]. $F_{YFS}$ is well-known from Ref. [5] and $C_A$ are the respective color Casimir invariants. We find that, when incorporated into the standard DGLAP-CS evolution, these improved kernels lead to ~5% changes in the moments of the NS densities [8] so that these improvements are of both technical and phenomenological significance. One of us (BFLW) has also worked-out the effect of these improvements on the higher order exact results for the kernels in Refs. [20] –see Refs. [8] where we show that there is no contradiction between (9) and the latter exact results.

Finally, in preparation for cross checks on various exact results which we need for the residuals in (1) and (2), one of us (BFLW) applied (1) to the issue of Bloch-Nordsieck non-cancellation in ISR in massive QCD at $O(\alpha_s^2)$[21,22,23]. As explained in [24], one finds that the real emission in the contribution $A_{q\text{-}o}$ in the notation of Ref.[22] saturates the IR single pole whose partial cancellation by the respective virtual corrections leads to the failure of Bloch-Nordsieck cancellation. One then applies (1) to the uncancelled fraction, $F_{nbn}$, and finds that exponentiation removes this uncancelled part of the real emission, restoring the respective Bloch-Nordsieck cancellation—see Refs.[24] for the details. The practical effect is that one can use non-zero quark masses in ISR at $O(\alpha_s^n)$, $n\geq 2$, in calculating the residuals in (1) and (2) and thereby cross-check the considerable massless results related thereto in the literature.

**5.0 Final state of Hawking radiation**

The many possible applications of the new approach to the UV divergence problem in quantum general relativity afforded by (3-5) are under study. Here let us record an important application by one us (BFLW) concerning the final state of the Hawking process[3] for an originally very massive black hole. Using the results (3-5), one computes the resummed graviton propagator and finds that Newton's potential is corrected[9] to

$$\Phi_N(r) = -\frac{G_N M}{r}(1 - e^{-ar}) \qquad (10)$$

where $a \cong 0.210\, M_{Pl}$. This is consistent with the asymptotic safety results in Ref.[25] and implies that elementary particles of the Standard Model are not black holes. Moreover, when one joins (10) on to the result in Ref.[25] for evolution of an originally very massive black hole by the Hawking process, one finds, consistent with Ref. [26], that the Planck scale remnant found in Ref.[25] becomes, due to (10), accessible to our universe, so that it should decay into Planck scale cosmic rays at least in some cases. One of us (BFLW) has encouraged experimentalists to look for such phenomena. The detailed rate of such events is under study—it is not known at this time.